\documentclass{sigchi}

\usepackage{balance}  
\usepackage{graphics} 
\usepackage{txfonts}
\usepackage{times}    
\usepackage[pdftex]{hyperref}
\usepackage{color}
\usepackage{textcomp}
\usepackage{booktabs}
\usepackage{ccicons}
\usepackage{todonotes}
\usepackage{subfigure}
\usepackage{csquotes}

\makeatletter
\def\url@leostyle{%
  \@ifundefined{selectfont}{\def\UrlFont{\sf}}{\def\UrlFont{\small\bf\ttfamily}}}
\makeatother
\def\pprw{8.5in}
\def\pprh{11in}

\definecolor{linkColor}{RGB}{6,125,233}
\hypersetup{%
  pdftitle={SIGCHI Conference Proceedings Format},
  pdfauthor={LaTeX},
  pdfkeywords={SIGCHI, proceedings, archival format},
  bookmarksnumbered,
  pdfstartview={FitH},
  colorlinks,
  citecolor=black,
  filecolor=black,
  linkcolor=black,
  urlcolor=linkColor,
  breaklinks=true,
}

\begin{document}

\title{Characterising Behavior and Emotions on Social Media for Safety: Exploring Online Communication between Police and Citizens}

\numberofauthors{3}
\author{%
  \alignauthor{Niharika Sachdeva\\
    \affaddr{CERC, IIIT-Delhi}\\
    \affaddr{India}\\
    \email{niharikas@iiitd.ac.in}}\\
  \alignauthor{Ponnurangam Kumaraguru\\
    \affaddr{CERC, IIIT-Delhi}\\
    \affaddr{India}\\
    \email{pk@iiitd.ac.in}}\\
}

\maketitle

\begin{abstract}

Increased use of social media by police to connect with citizens has encouraged researchers to study different aspects of information exchange (e.g. type of information, credibility and propagation) during emergency and crisis situation. Research studies lack understanding of human behavior such as engagement, emotions and social interaction between citizen and police department on social media. Several social media studies explore and show technological implications of human behavioral aspects in various contexts such as workplace interaction and depression in young mothers. In this paper, we study online interactions between citizens and Indian police in context of day-to-day policing, including safety concerns, advisories, etc. Indian police departments use Facebook to issue advisories, send alerts and receive citizen complaints and suggestions regarding safety issues and day-to-day policing. We explore how citizens express their emotions and social support on Facebook. Our work discusses technological implications of behavioral aspects on social well being of citizens.

\end{abstract}


\category{H.5.3}{Group and Organization Interfaces}[Collaborative computing, Computer-supported cooperative work]


\section{Introduction}
Urban societies exhibit major consternation over safety issues such as crime and law and order situations in day-to-day life. These issues has detrimental affects on the psychological well being of the victimised citizens and society~\footnote{society or community will be used interchangeably} at large (direct or indirect victimization~\cite{lewis1981community}). Research suggests that interaction between police and citizen plays an important role in overall perception of community towards safety~\cite{Stephens:2011fk}. A positive interactions (e.g. addressing the complaints) with police help increase trust and generates feeling of safety in the citizens whereas a negative ones (e.g., misconduct by police) increase insecurity among citizens~\cite{Stephens:2011fk}. Police understanding of citizens' behavior is helpful to know the effectiveness of safety measures, improve the organization's plan to address safety issues and involve citizens for collective action to improve safety~\cite{lewis1981community}.



Research suggests that social media (due to its massive reach) is a preferred medium to support interaction between citizen and government organisations such as police~\cite{SocialMediaPoliceUK,hughes2014online}. Prior work explores the effectiveness of using social media during crisis and post--crisis situation such as hurricane and fires, in developed countries~\cite{Cobb:2014uq, Palen-L.-and-Vieweg-S.:2008kx, Semaan:2012fk}.These studies demonstrate critical role of social media in civic engagement and social support to address safety concerns~\cite{export:208580,SocialMediaPoliceUK,hughes2014online,Lopez:2013:CCD:2441776.2441851}. However, it is unclear if the vast knowledge of human behavior on OSN (Online Social Networks) in event-driven situations such as crises can be helpful for police to understand citizens' behavior in the context of day-to-day policing. Few studies reflect on the utility of social media in day-to-day communication to prevent crime~\cite{DBLP:conf/ecscw/SachdevaK15, sachdeva2015social}. These studies discuss the information exchanged to enable increased engagement. However, citizens' behavior and emotions as they interact with police to address safety concerns are largely unexplored. Our study a) explores the feasibility of social media use to characterise and quantify the behavior and psychological factors influencing police and citizens interactions to address crime and safety concerns in day-to-day life; and b) compares human behavioral aspects such as emotions and social interactions when police and citizens interact on social media. To the best of our knowledge, this is the first work that empirically evaluates the role of social media in gauging changes in police and citizens' behavior to address crime and safety concerns in day-to-day life. 

In this work, we study content generated on 85 official police organisations Facebook pages to more comprehensively describe the changing trends in human behavior while responding to safety issues. Our dataset consists of 47,474 citizens' wall posts on police social media pages and 85,408 police status update from 2010 untill 2015. We describe various measures that can help operationalise affective response and social interactions in Facebook content generated by citizens participating on social media. These measures include three dimensions: engagement, emotional affect and intensity, and social process languages (linguistic attributes of language e.g. pronouns etc.) Indian Police departments are exhibiting fast adoption of social media to increase their presence and frequently interacting with citizens. India has the second largest users of Facebook and extensively uses Facebook for collective action to address safety concerns~\cite{Nayak:2014oq}. The policing department in India has only 130 personnel per 100,000 citizens where as United Nation guidelines recommends 270 -- 280 police personnel per 100,000 citizens to address day-to-day life issues~\cite{Express-News-Service:2013kx}. This lack of personnel in Indian police department results in many under policed areas and increases crime related concerns. Police feels the need of citizens' support to address crime issues and have taken to social media to interact with citizens~\cite{Skogan:2008fk,Stephens:2011fk}. Given these complex and unique policing conditions in India, social media communication has emerged as an effective solution to address safety concerns in day-to-day life.

\section{Research Objective}
The objective of this work is to~\emph{explore how human behavior such as engagement, emotional response and social response manifest on social media discussions between police and citizens to address safety concerns and day-to-day policing needs}. To examine this research objective, we study quantity and quality (behavioral) of discussions between police and citizens on social media. We evaluate quantity of interaction as~\emph{how much do citizens and police engage on social media to discuss safety issues}. To study quality, we look at \emph{how citizens and police use social media to express their emotions} and \emph{how does linguistics and social process between police and citizens manifest on OSN}. Literature shows engagement, emotions and linguistics are important aspects that help understand the human behavior~\cite{export:208580, DeChoudhury:2013:PPC:2470654.2466447, Kramer:2010:UBM:1753326.1753369}. Our work builds upon these studies in a novel policing context. 

\textbf{Towards Smart Cyber Safety Web}\\
Our work contributes and builds upon the prior knowledge in the HCI community on using OSN (Online Social Networks popularly known social media) in analysing human behavior, OSN use by citizens, first responders, and organizations for effective safety collaboration~\cite{Cobb:2014uq, Shklovski-I.-Palen-L.-and-Sutton-J.:2008vn, Stoll:2012zr, Voida:2012ys}. We show that OSN contributes to recording and sensing behavior such as emotions, social support and engagement of citizens in discussing safety issues. Our work illustrates that studying engagement models and linguistics patterns of OSN content provides enough evidences about psychological state of citizens concerned about safety and crime in day-to-day life. Consequently, acting as a natural laboratory that helps understand affects of safety concerns and collective action on social well-being. We highlight how discussions started by police and citizen on social media platforms differ in expressions. HCI community is in a unique position to use psychological cues to provide technological support that helps understand citizen opinion and improve strategies of collective action for safer society through social media. 

\section{Related Work}


Police forces are one of the fastest growing organisation on social media~\cite{Denef-S.-Kaptein-N.-Bayerl-P.-S.:2011vn, Lexis-Nexis-Risk-Solutions.:2012uq}. Research shows that social media offers two fundamental advantages for police: a) it can assist in primary policing job such as crime investigations, intelligence and prevention, and b) it can provide an instant direct communication platform with the public~\cite{heverin2010twitter}. A plethora of studies explore the role of social media during direct communication in critical events such as crisis. These studies show critical role of OSN in providing real-time information and reducing the spread of misinformation~\cite{Mendoza-M.-Poblete-B.-and-Castillo-C.:2010qf, Qu-Y.-Wu-P.-and-Wang-X.:2009ve, Starbird-K.-and-Palen-L.:2011bh, Vieweg-S-Hughes-A-Starbird-K-and-Palen-L.:2010dq}. Inspite of its various advantages, social media presence may be convoluting for police organisations as its adoption may leave them off-guard to continuous citizen scrutiny and comments~\cite{heverin2010twitter}.

Prior work validates social media contribution in improving citizen participation, responsiveness, and openness to different government organizations~\cite{Mergel:2014:SMA:2612733.2612740, Wigand:2010vn}. Social media provides government an instrument to develop opportunity for collective action involving both government authorities and citizens. Collective action approach improves governments' relationship with citizens and increases trust citizens place in the government~\cite{Kavanaugh:2011:SMU:2037556.2037574}. Unlike other organizations, police represent the most omnipresent and ubiquitous body of a society. The competing demands of citizens lead to greater expectations that police must be ``leaner'' and ``do more with less'' than other organizations~\cite{jeans1993relationship}. These expectations result in unique technology needs for the police department.

Majority of the work that contributes towards developing technological solutions for police focuses on designing systems to satisfy individual requirements of the police. These requirements vary from management of law enforcement content to use of video surveillance~\cite{Chen:2003:CML:602421.602441, Tullio:2010:EAE:1753326.1753551, Kavanaugh:2011:SMU:2037556.2037574}. Such technologies often ignore the citizens' view and needs for handling crime. On the other end, studies which focus on citizens as users offer solutions for personal safety, report unsafe locations, and wearable gadgets to reduce fear~\cite{Blom:2010:FCR:1753326.1753602, satchell2011welcome}. Such research endeavours capitalise on the principle of providing information to individuals so that they can avoid crime. Literature shows such an approach to avoid crime increases fear and victimisation affect in the citizens over a period of time~\cite{Lewis:2012:ETS:2207676.2208595}. Criminology research suggests that collective action that involves citizen and police both can successfully help reduce crime related anxiety and fear~\cite{lewis1981community}. However, very few studies focus on designing and studying technology that promote collective discussions between police and citizens to address safety concerns of citizens~\cite{heverin2010twitter, lewis1981community}. These studies discuss topics of interest and type of information exchanged during discussions between police and citizens but give little insight about what these discussions reflect on the human behavior such as emotional intelligence, social interactivity, and sensitisation affect induced in the society by such collective approach technologies. These aspects of human behavior are important elements of successful collaboration between organisation and individuals striving to achieve common goals~\cite{Lewis:2012:ETS:2207676.2208595, sypher1988communication, Xu:2012:LFA:2207676.2208524}.

HCI community has shown increased enthusiasm in understanding social media as an instrument to analyse human behavior and affective trends such as depression, desensitisation to violence, and happiness in various walks of life~\cite{export:208580, DeChoudhury:2013:PPC:2470654.2466447, Kramer:2010:UBM:1753326.1753369}. These studies explore activity, emotional exchange and linguistic aspects of content posted on social media to analyse human behavior in various public events involving collective crises response~\cite{ export:208580, verma2011natural}. Choudhury et al. examine the three affects of desensitization -- negative affect, activation and dominance observed on Twitter to show desentization among citizens living in Mexican Cities during Narco Wars~\cite{export:208580}. Few analyse the social media to track behavior of large population, for instance, Kramer used the posts people make on Facebook to evaluate happiness in the society~\cite{Kramer:2010:UBM:1753326.1753369}. Another study captures the temporal attributes to study the affect of positive and negative affect communicated on social media and its relation with the diurnal and seasonal behavior cultures~\cite{golder2011diurnal}. Other research works use linguistic features to measure and predict suicidal or death distress related symptoms on social media~\cite{brubaker2012grief, burnap2015machine}. These studies show that measures such as engagement with others, emotional expressions and linguistics patterns present in the content expressed on social media act as an effective tool to understand human behavior. 

We build upon these studies and investigate role of engagement, emotional response and social response that manifest on social media discussions between police and citizens to address safety concerns and improve day-to-day policing. Prior research discusses use of social media by citizens and police for general coordination during a crisis and different communities come in existence on OSN during crises ~\cite{Gupta-A.-Joshi-A.-and-Kumaraguru-P:2012cr, Hughes-A.-L.-Palen-J.-Sutton-S.-Liu-and-S.-Vieweg.:2008nx}. This information is useful for police and other law and order agencies to take decisions. Prior work reveals use of online social technologies by citizens and police mostly in crises and emergency situations. Very few studies explore use of social media by police and citizens in the context of day-to-day policing~\cite{DBLP:conf/ecscw/SachdevaK15,sachdeva2015social}. Such interactions play a crucial role in understanding direct / indirect victimisation effects of crime and safety concerns on citizens. In this work, we address these research gaps and explore role of social media discussions between police and citizens in day-to-day life to study the affective and psychological responses of citizens embroiled with safety concerns. This work has implications for government authorities and police officials working to improve policing (collective action) landscape and reduce affects of victimisation such as fear and anxiety among citizens using social media as a tool with a psychological affect.

HCI research can help develop technologies that help authorities understand behavioral aspects of communication and collaboration between police and citizens for better policing. For this, clearer insight into the behavioral aspects of technological interactions between police and citizens is required. Our research expands on the existing knowledge of OSN for law and order situation by providing a focused study that begins to address the specified research gaps. This work augments the traditional methods (focussed on information type) of evaluating and quantifying police - citizens interactions on social media. We believe that the insights from our study will provide opportunities to develop better communication strategy for police and citizens.

\section{Methodology}
The provisions so far adopted for policing in India are mostly non-technology based. Realizing the potential of social media to involve citizens in policing activities, Indian police have made its presence on Facebook (most popular social media in India)~\cite{DBLP:conf/ecscw/SachdevaK15}. We started by identifying all the police departments of India on Facebook. A government website~\footnote{http://arunpol.nic.in/} provides a list of all police departments. We extracted this list to collate all police departments on social media. We found that police departments exist at different levels: city, state, and district level. Using this list, we were able to find 100 police departments on Facebook. Next, we verified if these departments were true police department pages or not. To this end, we manually checked if these accounts were linked to the official government website of the police department or had stated on their webpage that these were the official government organisation representative. After this cleaning, we were left with 85 police departments pages on Facebook. For the police departments that we were not able to find on social media even after persistent and due diligence; we assumed that these departments either did not exist on social media or cannot be identified easily. In both the cases, such pages would have minimal utility for public communication to involve citizens in policing programs. Such a selection process has been used in the prior HCI studies as well~\cite{hughes2014online}.

\subsection{Data Collection}
We collected data from Facebook pages of all 85 police departments and used different mechanisms to filter the required posts and comments for analysis. We collected data using Facebook Graph API~\footnote{https://developers.facebook.com/docs/graph-api} from the day these pages were created till April 20$^{th}$ 2015. To study how discussions manifest on social media between police and citizens, we  collected all posts (wall posts\footnote{Content posted by citizen on police page.} and status updates\footnote{Content posted by police on its page.}) from these pages. In total, we collected 47,474 wall posts (i.e. made by citizens) and 85,408 status update (i.e. made by police) on these pages. Further, to understand discussions on these pages, we collected all comments, time of creation, message in comment etc. Our dataset had 46,845 police posts where we found atleast one comment and 24,984 citizen posts where there was atleast one comment. The data included in this study consists of only public posts and not include private messages that people might have sent to police using Facebook. Our dataset consists of pages with an average ages of 3 years. Few pages were very new (few months old) and some were as old as 5 years.


\subsection{Data Categorisation}
Our research is focussed on understanding behavioral aspects of discussions between police and citizens on social media. To understand the role of police interaction with citizens through social media, we look at discussions where both police and citizen participate and where only citizen participate in the discussion. In our dataset, we consider  discussions threads started by both police and citizens, i.e., we look at both citizen posts and police posts. We now explain the terminology and notations used in the paper to address different types of discussion threads.  In the notations, subscript represents who participates in the discussion (P\&C: both police and citizen comment and C: only citizens comment) and text represents who starts the discussion (P: Police Posts and C: Citizen Posts). Based on the who started the discussion (who created the post) and who participate in the discussion (who comments on the posts), we categorised data in four classes: a) police posts where both police and citizens have left comments and are part of discussion (P$_{P\&C}$), b) police posts where only citizens have left comment (P$_{C}$), c) citizen posts where police and citizens both have left comments (C$_{P\&C}$) and d) citizen posts where only citizens have left comment and are part of discussion (C$_{C}$). Table~\ref{datadescription} shows number of posts in each category where there was atleast one comment. In result section, we will study different interaction parameters across these four categories. Before discussing these parameters, we describe in brief the context of different posts in various categories in the next sub-section.



\begin{table}[]
\small
\centering
\caption{Describes total number of posts (Total posts), post with atleast one comment (One Comment), posts where both police and citizens comment (P\&C) and posts where only citizens comment  }
\label{datadescription}
\begin{tabular}{@{}lllll@{}}
\toprule
&Total  Posts           & One comment &  P\&C & Only C        \\ \midrule
Police   & 85,408                              & 46,845                             & 5,519 (P$_{P\&C}$)                                    & 41,326 (P$_{C}$)\\
Citizens & 47,474                              & 24,984                             & 17,196 (C$_{P\&C}$)                                     & 7,788 (C$_{C}$)  \\ \bottomrule
\end{tabular}
\end{table}

 \subsection{Understanding the Policing Context}
\begin{figure*}
\centering
\begin{tabular}{|l|r|}
\hline
\subfigure[P$_{P\&C}$]{{\includegraphics*[viewport= 0 70 770 430, angle=0, scale=0.31]{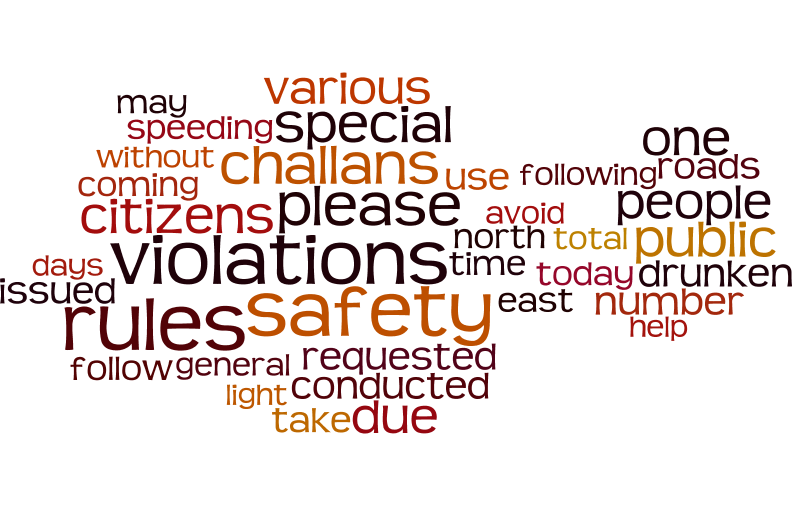}}} &
\subfigure[P$_{C}$]{{\includegraphics*[viewport= 0 60 790 440, angle=0, scale=0.31]{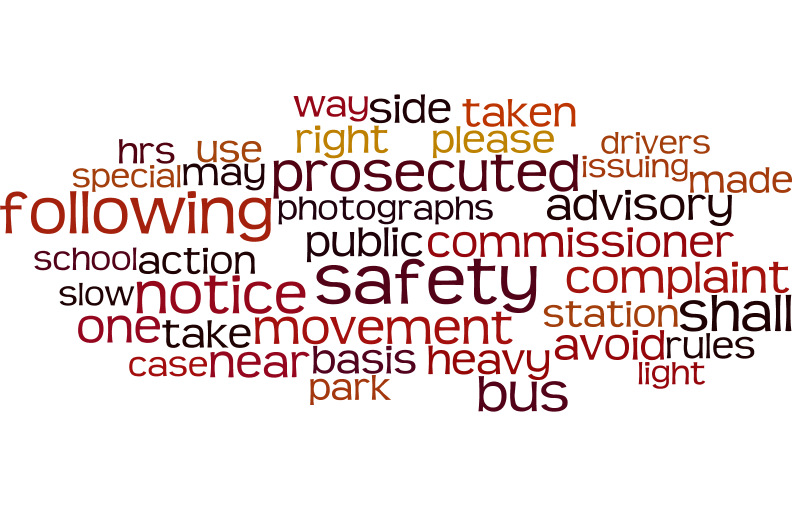}}} \\
\hline
\subfigure[C$_{P\&C}$]{{\includegraphics*[viewport= 0 60 790 440, angle=0, scale=0.31]{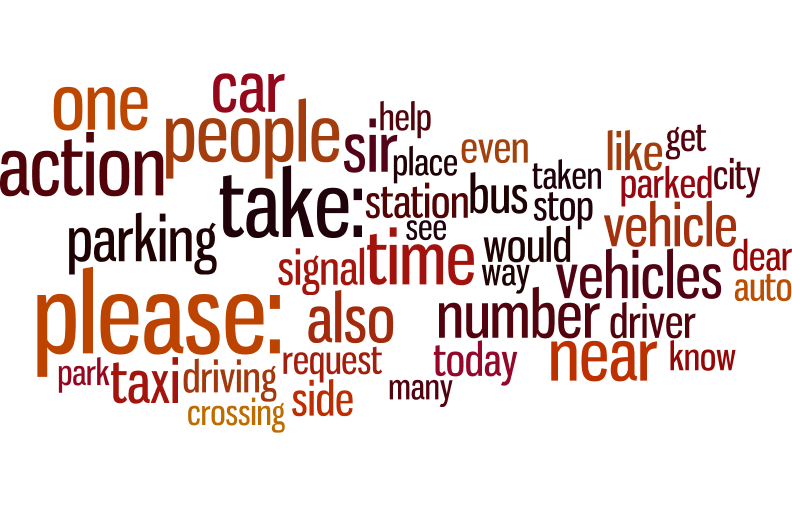}} }&
\subfigure[C$_{C}$]{{\includegraphics*[viewport= 0 60 790 440, angle=0, scale=0.31]{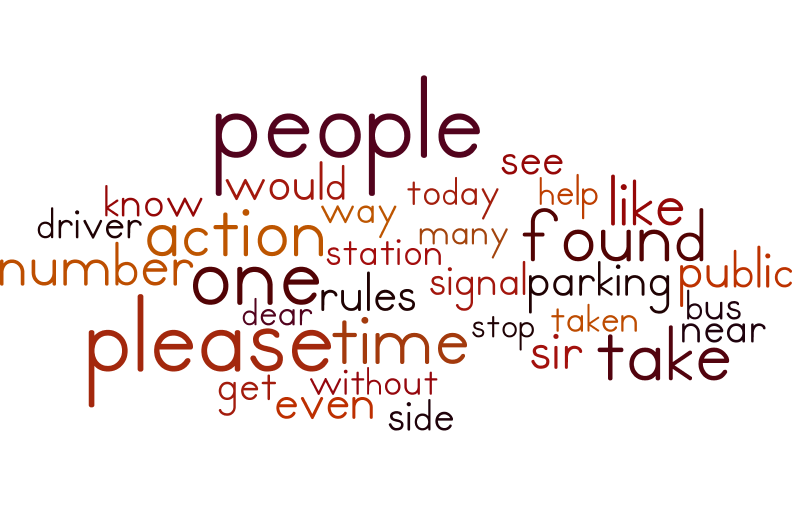}} }\\
\hline
\end{tabular}
\caption{Unigram Analysis: Relative strength of most frequent words across four data categories i.e. P$_{P\&C}$, P$_{C}$, C$_{P\&C}$ and C$_{C}$ discussions.}
\label{fig:TagCloud}
\end{figure*}

We did qualitative analysis followed by unigram analysis of text posted by police and citizen in their status updates and wall posts on Facebook pages. This analysis helped us set the context and  topic of discussion in different police and citizen posts where citizens and police leave comments.
We find that police posts included advisories, status of different cases being investigated, notice issued by police in public interests on various topics such as safety drives conducted, road safety, people prosecuted etc. Some of these posts also reflected actions such as challans (fine) issued to citizens based on complaints made by other citizens on Facebook page. These findings were further confirmed through unigram analysis, top words in the police posts were safety, citizens, people, notice, issued (See Figure~\ref{fig:TagCloud}). 


Next, within police posts, we compared P$_{P\&C}$ and P$_{P}$ discussions. In P$_{P\&C}$ discussions, we found that posts were broadly regarding general safety of public, e.g., in a post police asked citizens to share with their friends that snatching of objects is a big menace and remarked \emph{``Snatching is big menace in the cities especially in the colonies and during early hours, afternoons and nights. Although this menace cannot be eradicated totally, the vulnerability can be minimized, If you take the following precautions.....''} In another post police requested citizens to be careful while driving in rains, \emph{``Dear Car owners and Bike users , Please be careful while driving on a rainy day.....Do not splash the dirty pooled water on pedestrians just for fun... go slow and respect those who are passing by.''} In Figure~\ref{fig:TagCloud} (a), P$_{P\&C}$ posts show higher occurrence of words like safety, rules, people, citizens, requested and please showing that P$_{P\&C}$ posts advisories requesting citizens for specific actions or following rules.

In P$_{C}$ discussions (police posts where only citizens participate in discussions), we found that these posts were mostly regarding actions taken by police such as individuals prosecuted, notices issued etc. For instance, in a post police informed citizens that \emph{``29584 persons were prosecuted by Delhi Traffic Police for drunken driving during the year 2014. These drivers are requested not to drink \& drive again.....''} In another post, police stated that, \emph{``The following vehicles/owners have been prosecuted by issuing notice on the basis of photographs on dated 18/04/2013.
Vehicle no. - notice no.
DLXXXX0003-17XXXX41.......} In Figure~\ref{fig:TagCloud} (b), P$_{C}$ discussion's posts more words related to progress made by police department e.g. people prosecuted and action taken reports.

In the citizen posts, most posts requested police to take action on their complaints. The unigram analysis shows high occurrence of words please, take, action and sir etc. (See Figure~\ref{fig:TagCloud} (c) and (d)). On comparing citizen posts in C$_{P\&C}$ and C$_{C}$ discussions, we found that C$_{C}$ category had higher occurrence of posts having general reference to people and something citizens witnessed. For instance, a citizen said, \emph{``their so many foreigners whose visa is expired, other countries take serious action,why not in india, these people are mostly found in Yelahanka and around, some of them may be smugglers, terrorists, some of them spy working for enemy countries / terrorist org''}. In another post, a citizen said \emph{Many people's hard earned money are getting robbed without there knowledge. Guys please be careful when you go to petrol bunk, keep an Hawk eye on pump reading machine and verify Zero(0) before filling the tank and verify the exact amount filled before paying the cash.}

To summarize, police posts mostly consists of advisories and notice issued to citizens. In P$_{P\&C}$ discussions,  we found that most posts were regarding general safety of public where as in P$_{C}$ discussions most posts were regarding action and prosecution performed. In citizen posts, we find that most of the posts were complaints to the police. In C$_{P\&C}$, most posts were individual complaint where as in C$_{C}$ there was higher number of posts showing general public concerns.
\section{Results}
In this section, we study the Facebook posts and discussion threads between police and citizen to characterise and quantify the behaviour of police and citizens' interaction. We identify three measures to quantify this behavioral aspect of interaction between police and citizens; these measures include \emph{engagement}, \emph{emotional exchange}, and \emph{social interaction}. Our analysis is based on the literature quantifying human behaviour through social media in various context including depression, maternal health etc~\cite{export:208580,DeChoudhury:2013:PPC:2470654.2466447,de2014characterizing}.  We compare the behavioral measures on all four categories of discussions: P$_{P\&C}$, P$_{C}$, C$_{P\&C}$, C$_{C}$ (as discussed in last section) to understand how these aspects contribute and manifest themselves during discussions between police and citizens. 
We first describe our method of studying these measures followed by the findings. 
\vspace{10mm}
\subsection{Engagement}
In context of policing, engaging with citizens is an important aspect of the problem solving process~\cite{peak2002community}. Higher engagement with police during problem solving process may increase citizen trust and reduce differences in expectation between police and citizens~\cite{Skogan:2008fk}. In this sub-section, we discuss engagement as a measure of interaction. Volume of content generated is widely used as a measure of engagement~\cite{export:208580,DeChoudhury:2013:PPC:2470654.2466447}. In our work, we define two measures of volume -- a) number of content generators who participate in discussions~\footnote{Just the number of posts may not be useful in finding the engagement, as it will not capture the interaction i.e. leave a comment on a post.} and b) amount of content generated during these interaction (likes and comments). Together, we refer these measures of volume as engagement.


\subsubsection{Content generators}
In the citizen posts on police pages,  we find that number of people who comment on citizen posts where only citizens are part of discussion (C$_{C}$) is 12,630 which is 26\% lower than C$_{P\&C}$ discussions (where both police and citizens participate) (See Table~\ref{UGC}). This could be because the police posts are of use and address concerns of wider population. As we discussed in previous section, these are mostly advisories and actions taken by police whereas citizens' posts include individual complaints which may not relate to larger audience on police pages. It could therefore be that number of citizens participating in citizen posts are less than those participating in police posts. Prior work shows that social capital of organisations and individuals increases as more number of people contribute to content generation~\cite{gil2012social}. We find that C$_{P\&C}$ discussions contribute more towards social capital building (higher number content generators) than posts done by citizens where only citizens are part of discussions (See Table~\ref{UGC}).

To understand the contribution of citizens towards content generation process, we calculate number of  comments from citizens on each citizen post. We find that citizen comments are higher in C$_{P\&C}$  discussions when compared to C$_{C}$ (where only citizens comment). We calculate entropy (variation) in the number of comments per citizens in citizen posts. Entropy or variation in number of comments by each citizen is lower in C$_{C}$ discussion threads as compared to C$_{P\&C}$ discussions. Lower entropy value signifies that large number of comments are posted by small number of citizens. This behavior has been shown in other online social media studies, for instance, content generated on Twitter\cite{de2014characterizing,ghosh2011entropy}.


\begin{table}[]
\centering
\caption{Content Generators in discussions and Entropy change each users activity to comment (Ent/User). Only citizens shows content generators excluding police departments}
\label{UGC}
\begin{tabular}{@{}lllll@{}}
\toprule
&P$_{P\&C}$                 & P$_{C}$         & C$_{P\&C}$         & C$_{C}$                   \\ \midrule
Content Generators & 55,028       & 1,79,176      & 17,124       & 12,630       \\
Only citizens & 54,982		& 1,79,176	& 17,081 &12,630\\
Entropy             & 4.39 & 4.96 & 3.23 & 3.60 \\ \bottomrule
\end{tabular}
\end{table}

In posts made by police, we find that number of people who comment are higher in the P$_{C}$ (only citizens are part of discussion) than P$_{P\&C}$ discussions. Qualitative analysis shows that these posts ( P$_{C}$ discussions) are mainly general advisories and safety tips from police departments. These posts tend to get large number of comments by citizens where citizens thank police for sharing information with citizens. Following sample posts show example of posts from P$_{C}$ discussion (where only citizen participate).
\begin{displayquote}
\emph{How to obtain NOC from Traffic, Police Certificate, Duplicate RC?\\
Find the procedure Step by Step at the following link.\\
http://hyderabad.trafficpolice.co.in/?/noc-from-traffic-pol?}
\end{displayquote}

Next, to understand the contribution of individual citizens towards content generation process in police posts, we calculate entropy (variation) in the number of comments per citizen. We find that entropy is higher in P$_{C}$ as compared to P$_{P\&C}$. Higher entropy in P$_{C}$ discussions shows that there is less volatility in the content generators and citizens participation is more balanced in P$_{C}$ discussions (See Table~\ref{UGC}). 


\subsubsection{Content generated}
We analyse number of comments and likes on posts to understand which category of discussions get maximum interest and support. In citizen posts on police pages, we find that C$_{P\&C}$ discussion get less number of likes (9.48\% lower) and comments (29.75\% lower) compared to C$_{C}$  discussions (See Table~\ref{comm}). We find that these posts are complaints of individuals where they lost a vehicle, mobile phone, etc. or faced some problem in terms of FIR (First information Report) filed, complaints, etc. In most posts, police suggests an appropriate action and discussion closes early, resulting in less comments and likes by others. A sample post below illustrates this behaviour:

\begin{displayquote}
\emph{Citizen Post: Me and my family are getting unwanted call from the given number 83XXX93280. Specially he is misbehaving with a female member.
My Number is - 8VVVV936432
Pundrik CCCCC}

\emph{Police reply: Dear Pundrik CCCCC, Please visit at your nearest Police Station and lodge a complaint with details and they will assist you in this regard.. Thank you}
\end{displayquote}

\begin{table}[]
\small
\centering
\caption{Number of comments (Comm.) and likes on citizen posts and police posts. We used Mann Whitney U Test and found statistical differences. Tables shows Z values. ** represents p$<$0.01 and * represents p$<$0.05 }
\label{comm}
\begin{tabular}{@{}lllllll@{}}
\toprule
 &       & Comments      &   &        & Likes        &        \\ \midrule
         & Avg   & Std   & Med & Avg    & Std     & Med \\
P$_{P\&C}$      & 19.68 & 86.17 & 7      & 114.71 & 805.55  & 22     \\
P$_{C}$     & 9.88  & 74.92 & 3      & 88.05  & 1025.53 & 22     \\ \midrule
Z		&-48.25** & &  &-1.19 &&\\ \midrule
C$_{P\&C}$      & 3.34  & 19.19 & 2      & 9.4    & 253.85  & 2      \\
C$_{C}$     & 3.69  & 13.79 & 2      & 13.38  & 201.57  & 3      \\ \midrule
Z &-2.275**&&&-19.54*&&\\\bottomrule
\end{tabular}
\end{table}

We find that police posts where police and citizens both are part of discussion (P$_{P\&C}$) get more number of likes and comments than police posts where only citizens comment (P$_{C}$). As posts in P$_{P\&C}$ discussions are mainly advisories where police request citizens for some action and share information with them, these may generate more discussions among police and citizens. However, in P$_{C}$ discussions, posts are mainly notice issued or people prosecuted which may result in less discussions in the community. 

Previous research shows likes and comments contribute to the social support embedded in one's social network~\cite{de2014characterizing}. We observe, two contrasting aspects of engagement on police and citizen posts. In police posts, as P$_{P\&C}$ discussions get higher number of likes and comments than P$_{C}$. This shows that discussions on police posts where both police and citizens participate may contribute more to the social support of police on social network. Contrasting to police posts, we find that in C$_{P\&C}$ discussions (where both police and citizen participate) get less number of likes and comments whereas C$_{C}$ discussion threads get higher number of likes and comments. When police does not answer citizens' posts, number of likes are higher, as a result because other citizens participating on social media may show higher social support for the citizen who leaves a post on police page, but does not get a reply.


 
\subsection{Emotional Exchange}
Individual safety and threat are associated with high emotional activity. Research shows that interaction between police and citizens can reduce anxiety and apprehension~\cite{EmotionalAffect}. We analyse, if increased police interaction with citizen reduces such emotions and negative affect in discussions on police and citizen posts. To this end, we measure the~\emph{emotional level and emotional intensity} expressed by citizens and police on content posted in police pages.


\subsection{Emotional Affect}
In citizen posts, we find that emotional affect is 36.23\% higher in C$_{P\&C}$ discussions than C$_{C}$. We find statistically significant difference (See Table~\ref{NAvalues}) between the emotional affect observed in C$_{P\&C}$ and C$_{C}$ discussion threads. Emotional Affect captures the broader emotional state in content generated. We study two level of emotional affects: Negative (sad, angry, etc.) and Positive (happy, elated, etc.). We measure these affects using Linguistic Inquiry and Word Count (LIWC), a physcho-linguistic tool.~\footnote{http://liwc.wpengine.com/how-it-works/} Previous studies validate the use of LIWC in determining the emotional affects expressed in social media content~\cite{export:208580,DeChoudhury:2013:PPC:2470654.2466447}. 


\subsubsection{Negative Affect}
Negative Affect analysis shows citizens express negative emotions both in C$_{P\&C}$ and C$_{C}$ posts. However, in C$_{P\&C}$ discussions negative affect is 17.23\% higher than in C$_{C}$ discussions. We further analyse the distribution of anger, anxiety, and sadness in the negative sentiment expressed in  comments on citizen posts. We find that in C$_{P\&C}$ discussions, dominant contributor to negative affect is \emph{anger} followed by \emph{sadness} and \emph{anxiety} (See Table~\ref{NAvalues}). In C$_{C}$ discussions, we find that the dominant contributor to negative state is \emph{anger}, followed by \emph{anxiety} and \emph{sadness} (See Table~\ref{NAvalues}). 

We find that anxiety is significantly higher (177.78\%) in the C$_{C}$ discussion than C$_{P\&C}$. Higher anxiety in C$_{C}$ discussions indicates that discussions where only citizens comment, show higher levels \emph{nervousness} and \emph{worry} (expressions of anxiety) than the citizens discussions where police also participates. A reason for reduced anxiety in the discussions where police participate could be that frequent response from police may make citizen feel that the police is available to help and holds itself accountable to citizens. Previous work shows that this feeling of accountability may help reduce anxiety~\cite{EmotionalAffect,Lewis:2012:ETS:2207676.2208595} however sadness and anger remain because of the loss (such as theft) and victimization. Qualitative analysis of the posts confirm our findings; a citizen commented that:

\begin{displayquote}
\emph{I am just \textbf{worried} if Hyderabad Traffic Police [HTP] makes things worse like always and create more chaos. Frankly speaking ..... it's the lower income group or the people who are not aware using high beams. Try to educate people on road by campaigns at traffic signals, etc.}
\end{displayquote}

\begin{table}[]
\small
\centering
\caption{Values of Emotional Affect averaged across different citizen posts where citizens and police both participate (C$_{P\&C}$) in discussion and where only citizens participate in discussion (C$_{C}$). We used Mann Whitney U Test and found statistical differences. Tables shows Z values. ** represents p$<$0.01. PA = Positive Affect, NA = Negative Affect, Anx= Anxiety.}
\label{NAvalues}
\begin{tabular}{@{}llllllll@{}}
\toprule
&       &   C$_{P\&C}$     & &      &   C$_{C}$      &   &     \\ \midrule
                          & Avg   & Std   & Median                      & Avg  & Std    & Median & Z\\
Affect**                    & 10.04 & 11.3  & 7.47                        & 7.37 & 12.48  & 4.35  & -38.39\\
PA**                        & 7.57  & 11.28 & 4.76                        & 5.52 & 12.06 & 2.08 &  -37.46 \\
NA**                        & 2.11  & 3.04  & 0.67                        & 1.80  & 4.00 & 0.00    & -14.41 \\
Anx**                       & 0.09  & 0.76  & 0.00                           & 0.25 & 2.25  & 0.00   & -10.29  \\
Anger**                    & 0.60   & 1.69  & 0.00                           & 0.51 & 2.14  & 0.00    &  -3.52\\
Sad**                       & 0.12  & 0.67  & 0.00                           & 0.20 & 1.26 & 0.00  &  -8.03   \\ \bottomrule
\end{tabular}
\end{table}

We find similar anxiety trend in discussions on police posts. Anxiety is lower (64\%) in the discussions where police also participate (P$_{P\&C}$) than where only citizens participate in discussion~(See Table~\ref{PNAvalues}). 

On analysing positive affect on Facebook police pages, we find that C$_{P\&C}$ discussions (where police also participates) share 37.13\% higher positive sentiment than in the discussions where only citizens participate in discussions (C$_{C}$). Similar to previous studies, this shows that community discussions where police also participates can be used to transfuse higher positive sentiment among the citizens which are otherwise considered as coercive arm of the state~\cite{Lewis:2012:ETS:2207676.2208595}. 

\begin{table}[]
\small
\centering
\caption{Values of Emotional Affect averaged across different police posts where citizens and police both participate (P$_{P\&C}$) in discussion and where only citizens participate in discussion (P$_{C}$). We used Mann Whitney U Test and found statistical differences. Tables shows Z values. ** represents p$<$0.01. PA = Positive Affect, NA = Negative Affect, Anx= Anxiety.}
\label{PNAvalues}
\begin{tabular}{llllllll}
\toprule
 &      & P$_{P\&C}$     & &      &   P$_{C}$      &   &     \\ \midrule
                         & Avg  & Std  & Median                      & Avg  & Std   & Median & Z\\
Affect**                   & 8.30  & 7.38 & 6.49                        & 8.62 & 12.24 & 1.66  & -11.32  \\
PA**                       & 6.68 & 7.42 & 4.49                        & 7.02 & 12.22 & 1.78 & -12.99   \\
NA**                       & 1.57 & 1.9  & 1.28                        & 1.57 & 3.11  & 1.94 &-9.43  \\
Anx**                      & 0.19 & 0.5  & 0.00                           & 0.21 & 0.83  & 1.52  &-6.64 \\
Anger**                    & 0.50  & 0.89 & 0.00                           & 0.49 & 1.78  & 1.18 & -11.40   \\
Sad**                      & 0.16 & 0.50  & 0.00                           & 0.21 & 1.37  & 0.89 & -4.71 \\\bottomrule
\end{tabular}
\end{table}

We show that citizen posts exhibit higher emotional affect than police posts. In both C$_{P\&C}$  and P$_{P\&C}$ where both police and citizens participate, we find higher negative affect than posts where only citizens participate C$_{C}$  and P$_{C}$. We find that C$_{P\&C}$  and P$_{P\&C}$  discussions have less expression of anxiety and sadness than the discussions where only citizens are part of discussions (C$_{C}$  and P$_{C}$).


\subsubsection{Emotional Intensity}
We use two measures -- Arousal and Valence expressed in comments posted to measure the intensity of emotions involved on these pages. Research in the field of psychology shows that \emph{higher arousal tends to make people want to talk and communicate more. Hence people talk more when they are ``joyful'' and less when they are just ``contented.''} Similar effect can be seen with negative affect; infuriated is a higher arousal state whereas frustrated is lower. Valence is used to measure pleasure. To calculate these measures, we used ANEW lexicon implementation that helps to assess emotional intensities such as \emph{arousal} and \emph{valence} for a set of verbal terms of English Language~\footnote{http://www.csc.ncsu.edu/faculty/healey/maa-16/text/}. This has been used in previous works in the domain of analysing human behaviour such as depression and affective workplace presence~\cite{export:208580,DeChoudhury:2013:PPC:2470654.2466447,Xu:2012:LFA:2207676.2208524}. 

In our dataset, we find that C$_{P\&C}$ discussions show 12\% higher arousal than C$_{C}$ discussions (See Table~\ref{Citizen-EI}). We also find that for C$_{P\&C}$ discussions, there exists a stronger correlation between arousal and negative affect (r = 0.08) than arousal and positive affect (r = -0.04). This shows that arousal has more significant role to play when citizens are in negative affect state. We found similar trends in the discussions on police posts. Though these correlations are weak but they indicate the direction of relation between the variables. Posts in which both police and citizens participate in discussion show higher arousal (11\%) than discussions only among citizens (See Table~\ref{PCitizen-EI}).

We further observe that valence is 13\% higher in C$_{P\&C}$ discussions where police and citizens both participate than in C$_{C}$ discussions where only citizens participate (See Table~\ref{Citizen-EI}). On analysing the correlation of valence with positive and negative affect, we find that like arousal, valence is more strongly correlated with negative affect (r = 0.10) than with positive affect (r = -0.03). Though these correlations are weak but they indicate the direction of relation between the variables.This shows that valence plays a more significant role when the discussion between citizens \& police are in negative affect state, marking higher intensity in anger and fear.  


 

\begin{table}[]
\small
\centering
\caption{Values of Emotional Intensity averaged across different citizen posts where citizens and police both participate (C$_{P\&C}$) in discussion and where only citizens participate in discussion (C$_{C}$). We used Mann Whitney U Test and found statistical differences. Tables shows Z values. ** represents p$<$0.01.}
\label{Citizen-EI}
\begin{tabular}{llllllll}
\toprule
&&C$_{P\&C}$     &      &      & C$_{C}$   &      &              \\\midrule
        & Avg  & Std  & Median & Avg  & Std  & Median &Z \\
Arousal** & 4.40  & 1.74 & 5.01   & 3.90  & 2.16 & 4.66&  -11.74 \\
Valence** & 4.25 & 1.67 & 4.84   & 3.74 & 2.09 & 4.41 & -14.47\\ \bottomrule
\end{tabular}
\end{table}

\begin{table}[]
\small
\centering
\caption{Values of Emotional Intensity averaged across different police posts where citizens and police both participate (P$_{P\&C}$) in discussion and where only citizens participate in discussion (P$_{C}$). We used Mann Whitney U Test and found statistical differences. Tables shows Z values. ** represents p$<$0.01.}
\label{PCitizen-EI}
\begin{tabular}{@{}llllllll@{}}
\toprule
 &      &P$_{P\&C}$      &  &      & P$_{C}$     &   &     \\ \midrule
                         & Avg  & Std  & Median                      & Avg  & Std  & Median & Z \\
Arousal**                  & 4.19 & 1.23 & 4.35                        & 3.75 & 2.00    & 4.15 & -8.43 \\
Valence**                  & 4.13 & 1.27 & 4.25                        & 3.61 & 1.94 & 4.01  & -13.71 \\ \bottomrule
\end{tabular}
\end{table}

On analysing emotional affect and intensity in this sub-section, we find that in posts where citizens and police both participate in discussion are marked by higher negative affect and arousal than posts where only citizens participate in discussions. Also, in all posts (police and citizens), we find that negative affect is strongly correlated with arousal and valence showing that citizens on these pages may be more anxious to communicate their negative emotions. 

 
\subsection{Social Interactions}
Research shows that emphasis on social themes and efforts while connecting with other individuals causes stronger social and collaborative relationships among groups trying to achieve common objectives~\cite{Shami:2015:IEE:2702123.2702445}. Collaborative actions in the context of policing can increase social cohesion and reduce safety concerns among citizens~\cite{Lewis:2012:ETS:2207676.2208595}. To measure the social behaviour expressed on police pages, we study linguistics aspects such as mentioning family and friends and use of personal pronouns in the discussions on citizen and police posts. These help us compare if members socially engage during discussions on Facebook or restrict communication to sharing content about themselves to police. We use LIWC repository to measure linguistic aspects similar to other previous studies\cite{de2014characterizing,DeChoudhury:2013:PPC:2470654.2466447}.  



We find that in citizen posts, use of impersonal pronouns is higher in C$_{C}$ discussions whereas personal pronouns use is higher in C$_{P\&C}$ discussions. We find that C$_{C}$ discussion show higher  (82\% more) use of 1st person personal singular pronouns (e.g. \emph{I, me}) than discussions in which both police and citizens participate. Also, presence of 3rd person personal singular pronouns (e.g. she and he) and 3rd person plural pronoun (e.g. they) is higher in discussions where only citizens participate than where both police and citizens participate (See Table~\ref{Pronoun}). This indicates that discussions among citizens are highly self focused and citizens mostly express their own concerns that they face with others. We also find that C$_{C}$ discussions have more mentions of family, friends, and other people  than C$_{P\&C}$ discussions (See Table~\ref{Pronoun}). However, collective decision words such as 1st person personal plural pronouns (e.g. we and us) are very few in these conversations.  Qualitative analysis confirms our observation. 

\begin{displayquote} 
\emph{\textbf{I have} lived in the UK and all the time \textbf{I have never heard} anyone honking. Honking is not required if you know how to drive, this is a good move and hope to see some serious implementation. Can anyone advise me where to complain if \textbf{I see anyone} who do not comply ??
}
\end{displayquote}


\begin{displayquote}
\emph{kindly give me surety that \textbf{I will be safe and my family will be safe} after [I] expose them [Police officers] on Facebook.}
\end{displayquote}

The occurrence of 2nd person pronoun words (e.g. you) is lower in C$_{C}$ discussions than C$_{P\&C}$ discussions. Literature shows that occurrence of 2nd person pronoun \emph{you} is higher when people show increased focus on others~\cite{pennebaker2003psychological}. This focus could be to advice others or to hold others accountable. When police and citizens both participate in discussions, they tend to indulge in direct references to each other in the conversation. This may result in increased use of 2nd person pronouns. 
Following comments from the discussions confirm our observation:\\
Citizen comments:
\begin{displayquote}
\emph{Can \textbf{you} given any idea about the area.}
\end{displayquote}
\begin{displayquote}
\emph{Thanx.. Request \textbf{you to please} see to it that the law protectors also abides by the rules.}
\end{displayquote}
Police comments:
\begin{displayquote}
\emph{We are initiating a drive in Janipur from next week. Conscientious citizens like \textbf{you should} support.}
\end{displayquote}
\begin{displayquote}
\emph{
\textbf{You} should always ask for a receipt when \textbf{your offence} is compounded by an officer. REFUSE to hand over money......}
\end{displayquote}


Social interaction in C$_{C}$ discussions show higher tentativeness by frequently mentioning words like should, would, could and higher discrepancy such as guess, maybe, perhaps. Following comments from the discussions confirm our observation:

\begin{displayquote}
\emph{\textbf{I guess} the law needs to change. This needs to be restricted to law enforcement vehicles during emergency-only. In the above case there was only the driver in the car}.
\end{displayquote}
\begin{displayquote}
\textbf{I guess} its as per the law. PTP please confirm the same. Thanks in advance
\end{displayquote}
\begin{displayquote}
@Sujeet: \textbf{I guess} stolen vehicles are generally reported and an FIR is filed. So if the traffic cops have to verify if the vehicle is stolen all they have to do is take a printout of the license plate number and if they spot such a vehicle on the road they can take due action.......
\end{displayquote}

\begin{displayquote}
\emph{Well that is true about not looking up for the government for everything but ...... If \textbf{perhaps} they are directed towards such jobs as eco-tourism or \textbf{perhaps} even the police department (with a change to the ridiculous uniform of course).....}
\end{displayquote}

\begin{table}[]
\small
\centering
\caption{Use of Pronouns and Social Process like family, friends and humans in C$_{P\&C}$  and C$_{C}$. We used Mann Whitney U Test and found statistical differences. Tables shows Z values. ** represents p$<$0.01. ppron = personal pronouns, i = 1st person singular pronouns, we = 1st person plural pronouns, you = 2nd person shehe = 3rd person singular pronouns, they = 3rd person plural pronoun, ipron = impersonal pronouns}
\label{Pronoun}
\begin{tabular}{@{}llllllll@{}}
\toprule
 &      &    C$_{P\&C}$  &  &      &  C$_{C}$     &  &      \\ \midrule
                          & avg  & std  & median                      & avg  & std  & median & Z\\
ppron**                     & 6.16 & 5.88 & 5.26                        & 4.49 & 5.6  & 3.33 & -24.49   \\
i**                        & 0.78 & 1.67 & 0.00                           & 1.42 & 3.27 & 0.00   & -16.02  \\
we**                       & 1.38 & 3.11 & 0.00                          & 0.69 & 2.13 & 0.00   &  -16.01 \\
you**                      & 3.29 & 4.7  & 1.79                        & 1.18 & 3.18 & 0.00   &  -45.85 \\
shehe**                    & 0.24 & 0.96 & 0.00                           & 0.35 & 1.58 & 0.00 &  -4.44   \\
they**                      & 0.46 & 1.28 & 0.00                           & 0.85 & 2.18 & 0.00 &   -14.20  \\
ipron**                     & 2.99 & 3.63 & 2.3                         & 3.73 & 5.11 & 2.47 & -6.03  \\
family**                    & 0.05  & 0.48 & 0.00                           & 0.19 & 1.51 & 0.00  & -9.49     \\
friend**                    & 0.03  & 0.26 & 0.00                           & 0.10  & 0.98 & 0.00  & -8.17   \\
humans**                    & 0.78  & 2.55 & 0.00                           & 1.21 & 3.25 & 0.00   & -12.91  \\ 
tentativeness**                    & 1.47 & 2.40  & 0.00                           & 1.74 & 3.2  & 0.00  & -3.98    \\
Discrepancy** 		&0.92	&1.82	&0.00	&1.34	&2.88	&0.00 & -2.69\\
\bottomrule
\end{tabular}
\end{table}

We find similar trends of pronoun use in police posts (See Table~\ref{PronounPolice}). P$_{C}$ discussions have higher mention of 1st person singular pronouns and less mention of 1st person plural pronouns, showing high self focus. However, contrary to citizen posts we find that mention of family and friends was less in these posts.

In this section, we see that social interaction in C$_{C}$ and P$_{C}$ discussions, show higher self focus (greater use of 1st person singular pronouns), tentativeness and discrepancy than discussions where both police and citizens participate.  C$_{C}$ and P$_{C}$ discussions also show less mention of collective efforts than posts where both police and citizens participate.

\section{Discussion}

Recent studies on social media show its utility as instrument to analyse human behavior and affective trends such as depression, desensitisation to violence, and happiness in various walks of life~\cite{export:208580, DeChoudhury:2013:PPC:2470654.2466447, Kramer:2010:UBM:1753326.1753369}. However, these studies discuss little about social media role in exploring behavioral aspects of police and citizen interaction. Police organisations need sound understanding of citizen behavior to avoid violent upheavals. Our work explores the viability of using content generated on Facebook police pages as an instrument to quantify and characterise human behavior as police and citizens interact on Facebook to address safety concerns and day-to-day policing needs. We show that citizens use public platforms like Facebook to express their emotions and discuss collective solutions for safety with police and other members of civic society at large. Our work highlights that Facebook can be used to record and sense behavior such as emotions, social support and engagement of citizens in discussing safety issues with police and among themselves. These behavioral aspects can act as marked evidences of psychological state of citizens embroiled with safety concerns. Social media pages where citizens express their concerns act as a natural laboratory to understand the effects of increased safety needs on social well-being of society.

\begin{table}[]
\small
\centering
\caption{Use of Pronouns and Social Process like family, friends and humans in P$_{P\&C}$ and P$_{C}$. We used Mann Whitney U Test and found statistical differences. Tables shows Z values. ** represents p$<$0.01. ppron = personal pronouns, i = 1st person singular pronouns, we = 1st person plural pronouns, you = 2nd person shehe = 3rd person singular pronouns, they = 3rd person plural pronoun, ipron = impersonal pronouns.}
\label{PronounPolice}
\begin{tabular}{@{}llllllll@{}}
\toprule
 &      &P$_{P\&C}$      & &      & P$_{C}$      & &       \\ \midrule
                         & avg  & std  & median                     & avg  & std  & median&Z \\
ppron**                    & 5.09 & 3.64 & 4.72                       & 4.01 & 4.27 & 3.7 & -20.96  \\
i**                        & 1.25 & 1.59 & 0.92                       & 1.39 & 2.5  & 0.63 & -5.51 \\
we**                       & 0.83 & 1.31 & 0.40                        & 0.69 & 1.59 & 0.00 &-15.35     \\
you**                      & 2.08 & 3.17 & 1.19                       & 1.08 & 2.47 & 0.00  & -32.86   \\
shehe**                   & 0.24 & 0.73 & 0.00                          & 0.19 & 1.28 & 0.00    & -8.91 \\
they**                    & 0.69 & 1.06 & 0.00                          & 0.65 & 1.29 & 0.00  &  -9.96  \\
ipron**                    & 3.61 & 2.63 & 3.62                       & 3.91 & 4.28 & 3.59& -1.33  \\
family** & 0.11 & 0.48 & 0.00      & 0.09 & 0.62 & 0.00   &  -8.77 \\
friend** & 0.08 & 0.42 & 0.00      & 0.06 & 0.50  & 0.00   &   -10.03\\
humans** & 1.42 & 1.98 & 1.00      & 1.62 & 4.19 & 0.60 & -10.42  \\ \bottomrule
\end{tabular}
\end{table}

\subsection{Theoretical and Design Implications}

Understanding how citizens react to increased crime or law and order issue is important for police departments and authorities to gauge violent upheaval. Findings of this work can help police departments build necessary mechanisms and socio-psychological theories to respond to public complaints. We see that in Facebook discussions where only citizen participate show higher self focus (greater use of first person pronoun singular), tentativeness (guessing) and discrepancy than posts where both police and citizens participate. These discussions indicate weaker collective efforts than posts where both police and citizens participate. Contrary to this, discussions where police and citizens both participate show higher occurrence of 1st person plural pronouns. Literature shows collective action against safety issues reduces fear of lawlessness. However, posts where only citizens participate in discussions show how citizens may use public platform to voice their opinion and engage with other citizens to address accountability concerns. This behavior has been observed in other community platforms also where police does not answer citizen complaints. These can help imbibe increased transparency among police and citizens. Civil reactions to law and order situations are often strong and police needs to be aware of volatility and vulnerability of emotions in citizens to improve public relations. Observing citizens posts may help police make Informed choices that can help police avoid citizen backlashes and prevent loss of legitimacy and trust among citizens. 


We find some unique aspects of public communication between police and citizens on Facebook. Discussions where citizens and police both participate are marked by higher negative affect and arousal than posts where only citizens participate in discussions. Prior research show that this combination of emotions signals high sensitization among citizens~\cite{gebotys1988news}. Public discussions on platforms like Facebook can help police study sensitization in the citizens. Research show strong emotions can motivate citizens to take certain risks and reach out to others using social media for help~\cite{export:208580}. As researchers and designers, we can help design technologies that highlight such unique needs of citizens and contribute to understand sensitization to safety concerns.

Further, in all posts (police and citizens), we find that valence is more strongly correlated with negative affect than positive affect. More strong correlation with negative affect shows that citizens may be more anxious to communicate their negative sentiments. Higher correlation with negative affect is aligned with the theory of mutual sympathy of negative emotions and shows that citizens try to educate other citizens and police more about negative incidences they encounter~\cite{smith2010theory}. We envision tools that measure such emotional needs of citizens using social media content and help police and others to interact with citizens in need of emotional help. Such systems can also act as early predictors for authorities and police to understand changing behavior among citizens.

We find that posts where both citizens and police participate do not reduce negative affect, however, these posts show reduced anxiety than posts where only citizens participate. This may indicate that police interaction may help address anxiety related to safety concerns of citizens. We envision that designers could develop technologies that help gauge changing emotions among citizens. Using the measures discussed in this work, technology can be built to identify negative affects on citizens. These technologies could help police to record the reactions of citizens and share these records with decision makers to help take timely measures and gain better insights about citizens' concerns. Though in this work, we see potential use of social media to develop various technologies, we understand that these technologies may not work as standalone solutions. Rather, we believe that these solutions may complement existing methods and become part of broader detection systems and awareness programs about citizen's psychological and social responses. Different police departments can works towards developing ``interaction measures'' to identify emotional upheaval and social cohesion in the citizen groups that police interacts with on social media. Police can use these measures to supplement existing data about well being of the society.

This work has implications for government authorities and police officials working to improve policing landscape. Currently, understanding citizen reactions is a challenge for police departments as they mostly depend upon administrative and crime reports to gauge citizen reactions. These reports are often produced at an interval of a year or after a major civil upheaval thus making most of these approaches largely retrospective. We find that social media content can complement these existing methods to study psychological affect of safety concerns on citizens by providing real time knowledge about citizen behavior and activity.

\section{Limitations and Future Directions}
This study provides insights on OSN use in understanding behavioral aspects of police and citizen interaction. There are some known limitations to this work and our findings best be interpreted with some caution. The results of this work indicates explanatory variables influencing online interaction between police and citizens but require further proof through surveys and interviews. between police and citizen interaction using these results, however, the study helps to understand psychological well being and evaluating police efforts to increase citizen interaction  As part of future work, we plan to interview police and citizen participating on social media to understand what makes them share or inhibit them from sharing their thoughts on social media, to measure how much can we depend on these findings to be representative of citizens' thought process. 

In our methodology, we consider the self-reported emotions and social process using tools such as LIWC and ANEW. However, we cannot make claims about how much are these expressed words representative of real psychological state of citizens. India has culturally rich heritage and different regional language, however, for formal and official communication Indian constitution recognises English as the official language and is most popularly used. Therefore the maximum content on these pages was in English. Thus, helping us to use standard methods used on English Language for analysis in our study. Other studies in India, have also used such methods~\cite{perez2014cross}.

We only study users from urban areas where OSN influence is high. It will be interesting to study a broader space as there may exists other factors such as demography, education, cultural background that may influence expression of human behavior on social media. It will be interesting to study how ground situation such as crime rates and law and order issues influence the nature of interaction between police and citizens. We leave these questions to be answered in the followup work. In this work, we explore interaction between police and citizens on Facebook, however, there are multiple other social networks which police is exploring such as Twitter and YouTube; It will be interesting to compare how our findings expand to other social networks.

\section{Acknowledgement}
We would like to thank TCS research for funding the project. Also, we would like to thank all the members of Cybersecurity Education and Research Center and Precog who have given us continued support throughout the project; special thanks to Prateek Dewan and Siddhartha Asthana.

\bibliographystyle{SIGCHI-Reference-Format}
\bibliography{OSMpolicing}


\begin{thebibliography}{00}


\ifx \showCODEN    \undefined \def \showCODEN     #1{\unskip}     \fi
\ifx \showDOI      \undefined \def \showDOI       #1{{\tt DOI:}\penalty0{#1}\ }
  \fi
\ifx \showISBNx    \undefined \def \showISBNx     #1{\unskip}     \fi
\ifx \showISBNxiii \undefined \def \showISBNxiii  #1{\unskip}     \fi
\ifx \showISSN     \undefined \def \showISSN      #1{\unskip}     \fi
\ifx \showLCCN     \undefined \def \showLCCN      #1{\unskip}     \fi
\ifx \shownote     \undefined \def \shownote      #1{#1}          \fi
\ifx \showarticletitle \undefined \def \showarticletitle #1{#1}   \fi
\ifx \showURL      \undefined \def \showURL       #1{#1}          \fi

\bibitem{Blom:2010:FCR:1753326.1753602}
{Jan Blom}, {Divya Viswanathan}, {Mirjana Spasojevic}, {Janet Go}, {Karthik
  Acharya}, {and} {Robert Ahonius}.
\newblock \showarticletitle{Fear and the City: Role of Mobile Services in
  Harnessing Safety and Security in Urban Use Contexts}. In {\em Proc. CHI
  2010}. ACM, 1841--1850.
\newblock
\showISBNx{978-1-60558-929-9}
\showURL{%
\url{http://doi.acm.org/10.1145/1753326.1753602}}


\bibitem{brubaker2012grief}
{Jed~R Brubaker}, {Funda Kivran-Swaine}, {Lee Taber}, {and} {Gillian~R Hayes}.
  2012.
\newblock \showarticletitle{Grief-Stricken in a Crowd: The Language of
  Bereavement and Distress in Social Media.}
\newblock


\bibitem{burnap2015machine}
{Pete Burnap}, {Walter Colombo}, {and} {Jonathan Scourfield}. 2015.
\newblock \showarticletitle{Machine classification and analysis of
  suicide-related communication on twitter}. In {\em Proceedings of the 26th
  ACM Conference on Hypertext \& Social Media}. ACM, 75--84.
\newblock


\bibitem{Chen:2003:CML:602421.602441}
{Hsinchun Chen}, {Daniel Zeng}, {Homa Atabakhsh}, {Wojciech Wyzga}, {and}
  {Jenny Schroeder}. 2003.
\newblock \showarticletitle{COPLINK: Managing Law Enforcement Data and
  Knowledge}.
\newblock {\em Commun. ACM\/} {46}, 1 (Jan. 2003), 28--34.
\newblock
\showISSN{0001-0782}
\showDOI{%
\url{http://dx.doi.org/10.1145/602421.602441}}


\bibitem{export:208580}
{Munmun~De Choudhury}, {Andres Monroy-Hernandez}, {and} {Gloria Mark}. 2014.
\newblock "Narco" Emotions: Affect and Desensitization in Social Media during
  the Mexican Drug War.
\newblock   (2014).
\newblock
\showURL{%
\url{http://research.microsoft.com/apps/pubs/default.aspx?id=208580}}


\bibitem{Cobb:2014uq}
{Camille Cobb}, {Ted McCarthy}, {Annuska Perkins}, {Ankitha Bharadwaj}, {Jared
  Comis}, {Brian Do}, {and} {Kate Starbird}. 2014.
\newblock \showarticletitle{{Designing for the Deluge: Understanding and
  Supporting the Disturbed, Collaborative Work of Crisis Volunteers}}.
\newblock {\em {CSCW}\/} (2014).
\newblock


\bibitem{DeChoudhury:2013:PPC:2470654.2466447}
{Munmun De~Choudhury}, {Scott Counts}, {and} {Eric Horvitz}. 2013.
\newblock \showarticletitle{Predicting Postpartum Changes in Emotion and
  Behavior via Social Media}. In {\em Proceedings of the SIGCHI Conference on
  Human Factors in Computing Systems} {\em (CHI '13)}. ACM, New York, NY, USA,
  3267--3276.
\newblock
\showISBNx{978-1-4503-1899-0}
\showDOI{%
\url{http://dx.doi.org/10.1145/2470654.2466447}}


\bibitem{de2014characterizing}
{Munmun De~Choudhury}, {Scott Counts}, {Eric~J Horvitz}, {and} {Aaron Hoff}.
  2014.
\newblock \showarticletitle{Characterizing and predicting postpartum depression
  from shared facebook data}. In {\em Proceedings of the 17th ACM conference on
  Computer supported cooperative work \& social computing}. ACM, 626--638.
\newblock


\bibitem{EmotionalAffect}
{Louis~M. Dekmar}. 2015.
\newblock Handling Citizen Complaints through Proactive Methodology.
\newblock
  \url{http://www.policechiefmagazine.org/magazine/index.cfm?fuseaction=display_arch&article_id=2052&issue_id=42010}.
    (Sept. 2015).
\newblock


\bibitem{SocialMediaPoliceUK}
{Sebastian Denef}, {Petra~S. Bayerl}, {and} {Nico Kaptein}.
\newblock \showarticletitle{{Social Media and the Police - Tweeting Practices
  of British Police Forces during the August 2011 Riots}}. In {\em Proc. CHI
  2013}. ACM, 3471 -- 3480.
\newblock


\bibitem{Denef-S.-Kaptein-N.-Bayerl-P.-S.:2011vn}
{{Denef, S., Kaptein, N., Bayerl, P. S.}} 2011.
\newblock \showarticletitle{{ICT Trends in European Policing. The COMPOSITE
  Project.}}
\newblock http://www.fit.fraunhofer.de/content/dam/fit/de/docume
  nts/composite\_d41.pdf,  (2011).
\newblock


\bibitem{Express-News-Service:2013kx}
{{Express News Service}}. 2013.
\newblock {Is community policing need of the hour?}
\newblock \urlstyle{same}
  \url{http://www.newindianexpress.com/states/karnataka/article1430481.ece}.
  (2013).
\newblock


\bibitem{gebotys1988news}
{Robert~J Gebotys}, {Julian~V Roberts}, {and} {Bikram DasGupta}. 1988.
\newblock \showarticletitle{News media use and public perceptions of crime
  seriousness}.
\newblock {\em Canadian J. Criminology\/}  {30} (1988), 3.
\newblock


\bibitem{ghosh2011entropy}
{Rumi Ghosh}, {Tawan Surachawala}, {and} {Kristina Lerman}. 2011.
\newblock \showarticletitle{Entropy-based classification of'retweeting'activity
  on twitter}.
\newblock {\em arXiv preprint arXiv:1106.0346\/} (2011).
\newblock


\bibitem{gil2012social}
{Homero Gil~de Z{\'u}{\~n}iga}, {Nakwon Jung}, {and} {Sebasti{\'a}n
  Valenzuela}. 2012.
\newblock \showarticletitle{Social media use for news and individuals' social
  capital, civic engagement and political participation}.
\newblock {\em Journal of Computer-Mediated Communication\/} {17}, 3 (2012),
  319--336.
\newblock


\bibitem{golder2011diurnal}
{Scott~A Golder} {and} {Michael~W Macy}. 2011.
\newblock \showarticletitle{Diurnal and seasonal mood vary with work, sleep,
  and daylength across diverse cultures}.
\newblock {\em Science\/} {333}, 6051 (2011), 1878--1881.
\newblock


\bibitem{Gupta-A.-Joshi-A.-and-Kumaraguru-P:2012cr}
{{Gupta, A., Joshi, A., and Kumaraguru, P}}. 2012.
\newblock \showarticletitle{{Identifying and Characterizing User Communities on
  Twitter during Crisis Events}}.
\newblock {\em Workshop on UMSocial, Co-located with CIKM\/} (2012).
\newblock


\bibitem{heverin2010twitter}
{Thomas Heverin} {and} {Lisl Zach}. 2010.
\newblock \showarticletitle{Twitter for city police department information
  sharing}.
\newblock {\em Proceedings of the American Society for Information Science and
  Technology\/} {47}, 1 (2010), 1--7.
\newblock


\bibitem{hughes2014online}
{Amanda~L Hughes}, {Lise~AA St~Denis}, {Leysia Palen}, {and} {Kenneth~M
  Anderson}.
\newblock \showarticletitle{Online public communications by police \& fire
  services during the 2012 Hurricane Sandy}. In {\em Proc. CHI 2014}. ACM,
  1505--1514.
\newblock


\bibitem{Hughes-A.-L.-Palen-J.-Sutton-S.-Liu-and-S.-Vieweg.:2008nx}
{{Hughes, A., L. Palen, J. Sutton, S. Liu, and S. Vieweg.}} 2008.
\newblock \showarticletitle{{``Site- Seeing'' in disaster: An examination of
  on-line social convergence}.}
\newblock {\em ISCRAM\/} (2008).
\newblock


\bibitem{jeans1993relationship}
{Dianne Jeans}. 1993.
\newblock \showarticletitle{{The Relationship Between Police and Other
  Government Agencies: Recent Changes in Perspective in Queensland}}.
\newblock


\bibitem{Kavanaugh:2011:SMU:2037556.2037574}
{Andrea Kavanaugh}, {Edward~A. Fox}, {Steven Sheetz}, {Seungwon Yang}, {Lin~Tzy
  Li}, {Travis Whalen}, {Donald Shoemaker}, {Paul Natsev}, {and} {Lexing Xie}.
  2011.
\newblock \showarticletitle{Social Media Use by Government: From the Routine to
  the Critical}. In {\em Proceedings of the Annual International Digital
  Government Research Conference: Digital Government Innovation in Challenging
  Times} {\em (dg.o '11)}. ACM, 121--130.
\newblock
\showISBNx{978-1-4503-0762-8}
\showDOI{%
\url{http://dx.doi.org/10.1145/2037556.2037574}}


\bibitem{Kramer:2010:UBM:1753326.1753369}
{Adam~D.I. Kramer}. 2010.
\newblock \showarticletitle{An Unobtrusive Behavioral Model of "Gross National
  Happiness"}. In {\em Proceedings of the SIGCHI Conference on Human Factors in
  Computing Systems} {\em (CHI '10)}. ACM, New York, NY, USA, 287--290.
\newblock
\showISBNx{978-1-60558-929-9}
\showDOI{%
\url{http://dx.doi.org/10.1145/1753326.1753369}}


\bibitem{lewis1981community}
{Dan~A Lewis} {and} {Greta Salem}. 1981.
\newblock \showarticletitle{Community crime prevention: An analysis of a
  developing strategy}.
\newblock {\em Crime \& Delinquency\/} {27}, 3 (1981), 405--421.
\newblock


\bibitem{Lewis:2012:ETS:2207676.2208595}
{Sheena Lewis} {and} {Dan~A. Lewis}.
\newblock \showarticletitle{Examining Technology That Supports Community
  Policing}. In {\em Proc. CHI 2012}. ACM, 1371--1380.
\newblock
\showISBNx{978-1-4503-1015-4}
\showDOI{%
\url{http://dx.doi.org/10.1145/2207676.2208595}}


\bibitem{Lexis-Nexis-Risk-Solutions.:2012uq}
{{Lexis Nexis Risk Solutions.}} 2012.
\newblock Survey of Law Enforcement Personnel and Their Use of Social Media in
  Investigations.
\newblock \url{www.lexisnexis.com/investigations}.   (2012).
\newblock


\bibitem{Lopez:2013:CCD:2441776.2441851}
{Claudia~A. L\'{o}pez} {and} {Brian~S. Butler}. 2013.
\newblock \showarticletitle{Consequences of Content Diversity for Online Public
  Spaces for Local Communities}. In {\em Proceedings of the 2013 Conference on
  Computer Supported Cooperative Work} {\em (CSCW '13)}. ACM, 673--682.
\newblock
\showISBNx{978-1-4503-1331-5}
\showDOI{%
\url{http://dx.doi.org/10.1145/2441776.2441851}}


\bibitem{Mendoza-M.-Poblete-B.-and-Castillo-C.:2010qf}
{{Mendoza, M., Poblete, B. and Castillo, C.}} 2010.
\newblock \showarticletitle{{Twitter under crisis: Can we trust what we RT?}}
\newblock {\em 1st Workshop on Social Media Analytics SOMA\/} (2010).
\newblock


\bibitem{Mergel:2014:SMA:2612733.2612740}
{Ines Mergel}. 2014.
\newblock \showarticletitle{Social Media Adoption: Toward a Representative,
  Responsive or Interactive Government?}. In {\em Proceedings of the Annual
  International Conference Digital Government Research} {\em (dg.o '14)}. ACM,
  163--170.
\newblock
\showISBNx{978-1-4503-2901-9}
\showDOI{%
\url{http://dx.doi.org/10.1145/2612733.2612740}}


\bibitem{Nayak:2014oq}
{Varun Nayak}. 2014.
\newblock {92 Million Facebook Users Makes India The Second Largest Country
  [STUDY]}.
\newblock \urlstyle{same}
  \url{http://www.dazeinfo.com/2014/01/07/facebook-inc-fb-india-demographic-users-2014}.
    (2014).
\newblock


\bibitem{Palen-L.-and-Vieweg-S.:2008kx}
{{Palen, L. and Vieweg, S.}} 2008.
\newblock \showarticletitle{{The Emergence of Online Widescale Interaction:
  Assiatance, Alliance and Retreat}}.
\newblock {\em {CSCW}\/} (2008).
\newblock


\bibitem{peak2002community}
{Kenneth~J Peak} {and} {Ronald~W Glensor}. 2002.
\newblock {\em Community policing and problem solving: Strategies and
  practices}.
\newblock Prentice Hall Upper Saddle River, NJ.
\newblock


\bibitem{pennebaker2003psychological}
{James~W Pennebaker}, {Matthias~R Mehl}, {and} {Kate~G Niederhoffer}. 2003.
\newblock \showarticletitle{Psychological aspects of natural language use: Our
  words, our selves}.
\newblock {\em Annual review of psychology\/} {54}, 1 (2003), 547--577.
\newblock


\bibitem{perez2014cross}
{Ver{\'o}nica P{\'e}rez-Rosas} {and} {Rada Mihalcea}.
\newblock \showarticletitle{Cross-cultural deception detection}.
\newblock


\bibitem{Qu-Y.-Wu-P.-and-Wang-X.:2009ve}
{{Qu, Y., Wu, P. and Wang, X.}} 2009.
\newblock \showarticletitle{{Online Community Response to Major Disaster: A
  Case Study of Tianya Forum in the 2008 China Earthquake.}}
\newblock {\em In Proc 42nd Hawaii Int'l Conf. on System Sciences\/} (2009).
\newblock


\bibitem{DBLP:conf/ecscw/SachdevaK15}
{Niharika Sachdeva} {and} {Ponnurangam Kumaraguru}. 2015a.
\newblock \showarticletitle{Online Social Networks and Police in India -
  Understanding the Perceptions, Behavior, Challenges}. In {\em {ECSCW} 2015:
  Proceedings of the 14th European Conference on Computer Supported Cooperative
  Work, 19-23 September 2015, Oslo, Norway}. 183--203.
\newblock
\showDOI{%
\url{http://dx.doi.org/10.1007/978-3-319-20499-4_10}}


\bibitem{sachdeva2015social}
{Niharika Sachdeva} {and} {Ponnurangam Kumaraguru}. 2015b.
\newblock \showarticletitle{Social networks for police and residents in India:
  exploring online communication for crime prevention}. In {\em Proceedings of
  the 16th Annual International Conference on Digital Government Research}.
  ACM, 256--265.
\newblock


\bibitem{satchell2011welcome}
{Christine Satchell} {and} {Marcus Foth}.
\newblock \showarticletitle{Welcome to the jungle: HCI after dark}. In {\em CHI
  2011 Extended Abstracts on Human Factors in Computing Systems}. ACM,
  753--762.
\newblock


\bibitem{Semaan:2012fk}
{Bryan Semaan} {and} {Gloria Mark}. 2012.
\newblock \showarticletitle{{`Facebooking' Towards Crisis Recovery and Beyond:
  Disruption as an Opportunity}}.
\newblock {\em {In Proc. CSCW}\/} (2012), 27--36.
\newblock


\bibitem{Shami:2015:IEE:2702123.2702445}
{N.~Sadat Shami}, {Michael Muller}, {Aditya Pal}, {Mikhil Masli}, {and} {Werner
  Geyer}. 2015.
\newblock \showarticletitle{Inferring Employee Engagement from Social Media}.
  In {\em Proceedings of the 33rd Annual ACM Conference on Human Factors in
  Computing Systems} {\em (CHI '15)}. ACM, New York, NY, USA, 3999--4008.
\newblock
\showISBNx{978-1-4503-3145-6}
\showDOI{%
\url{http://dx.doi.org/10.1145/2702123.2702445}}


\bibitem{Shklovski-I.-Palen-L.-and-Sutton-J.:2008vn}
{{Shklovski, I., Palen, L., and Sutton, J.}} 2008.
\newblock \showarticletitle{Finding Community through information and
  communication technology in disaster response}.
\newblock {\em {CSCW}\/} (2008).
\newblock


\bibitem{Skogan:2008fk}
{Wesley~G. Skogan}. 2008.
\newblock {\em An Overview of Community Policing: Origins, Concepts and
  Implementation}.
\newblock Number 43-57 in The Handbook of Knowledge-Based Policing: Current
  Conceptions and Future Directions. Wiley.
\newblock


\bibitem{smith2010theory}
{Adam Smith}. 2010.
\newblock {\em The theory of moral sentiments}.
\newblock Penguin.
\newblock


\bibitem{Starbird-K.-and-Palen-L.:2011bh}
{{Starbird, K. and Palen, L.}} 2011.
\newblock \showarticletitle{{``Voluntweeters'': Self- organizing by digital
  volunteers in times of crisis.}}
\newblock {\em In Proc. CHI\/} (2011), 1071--1080.
\newblock


\bibitem{Stephens:2011fk}
{Darrel~W. Stephens}, {Julia Hill}, {and} {Sheldon Greenberg}. 2011.
\newblock {Strategic Communication Practices: A Toolkit for Police Executives}.
\newblock {Community Oriented Policing Services, U.S. Department of Justice}.
  (Sept. 2011).
\newblock


\bibitem{Stoll:2012zr}
{Jennifer Stoll}, {kirsten Foot}, {and} {W.~Keith Edwards}. 2012.
\newblock \showarticletitle{{Between Us and Them: Building Connectedness Within
  Civic Networks}}.
\newblock {\em {In Proc. CSCW}\/} (2012), 237--240.
\newblock


\bibitem{sypher1988communication}
{Howard~E Sypher} {and} {Edward~Tory Higgins}. 1988.
\newblock {\em Communication, social cognition, and affect}.
\newblock Psychology Press.
\newblock


\bibitem{Tullio:2010:EAE:1753326.1753551}
{Joe Tullio}, {Elaine Huang}, {David Wheatley}, {Harry Zhang}, {Claudia
  Guerrero}, {and} {Amruta Tamdoo}.
\newblock \showarticletitle{Experience, Adjustment, and Engagement: The Role of
  Video in Law Enforcement} {\em (In Proc CHI 2010)}. ACM, 1505--1514.
\newblock
\showISBNx{978-1-60558-929-9}
\showDOI{%
\url{http://dx.doi.org/10.1145/1753326.1753551}}


\bibitem{verma2011natural}
{Sudha Verma}, {Sarah Vieweg}, {William~J Corvey}, {Leysia Palen}, {James~H
  Martin}, {Martha Palmer}, {Aaron Schram}, {and} {Kenneth~Mark Anderson}.
  2011.
\newblock \showarticletitle{Natural Language Processing to the Rescue?
  Extracting" Situational Awareness" Tweets During Mass Emergency.} Citeseer.
\newblock


\bibitem{Vieweg-S-Hughes-A-Starbird-K-and-Palen-L.:2010dq}
{{Vieweg, S., Hughes, A., Starbird, K., and Palen, L.}} 2010.
\newblock \showarticletitle{Micro-blogging during two natural hazards events:
  What Twitter May Contribute to Situational Awareness}.
\newblock {\em In Proc CHI\/} (2010), 1079--1088.
\newblock


\bibitem{Voida:2012ys}
{Amy Voida}, {Ellie Harmon}, {and} {Ban Al-Ani}.
\newblock \showarticletitle{{Bridging Between Organisation and the Public:
  Volunteer Coordinators' Uneasy Relationship with Social Computing}}. In {\em
  Proc. CHI 2012} {\em (ACM)}. 1967 -- 1976.
\newblock


\bibitem{Wigand:2010vn}
{F.~Dianne~Lux Wigand}. 2010.
\newblock \showarticletitle{Twitter Takes Wing in Government: Diffusion, Roles,
  and Management}.
\newblock {\em 11th Annual International Conference on Digital Government
  Research\/} (2010).
\newblock


\bibitem{Xu:2012:LFA:2207676.2208524}
{Anbang Xu}, {Jacob Biehl}, {Eleanor Rieffel}, {Thea Turner}, {and} {William
  van Melle}. 2012.
\newblock \showarticletitle{Learning How to Feel Again: Towards Affective
  Workplace Presence and Communication Technologies}. In {\em Proceedings of
  the SIGCHI Conference on Human Factors in Computing Systems} {\em (CHI '12)}.
  ACM, New York, NY, USA, 839--848.
\newblock
\showISBNx{978-1-4503-1015-4}
\showDOI{%
\url{http://dx.doi.org/10.1145/2207676.2208524}}


\end{thebibliography}
\balance

\end{document}